\documentclass[12pt]{article}
\pdfoutput=1
\usepackage{amsbsy}
\usepackage{amsfonts}
\usepackage{amsmath,amssymb}
\usepackage{bbold}
\usepackage{graphicx}

\usepackage[makeroom]{cancel}

\newcommand{\sect}[1]{\setcounter{equation}{0}\section{#1}}

\newcommand{\eq}{\begin{equation}}
\newcommand{\eqa}{\begin{eqnarray}}  
\newcommand{\en}{\end{equation}}
\newcommand{\ena}{\end{eqnarray}}
\newcommand{\enn}{\nonumber \end{equation}}

\def\sk{\vskip .4cm}
\def\noi{\noindent}

\def\al{\alpha}
\def\be{\beta}

\let \part\partial

\def\part{\partial}

\def\sk{\vskip .4cm}

\def\noi{\noindent}

\def\X0{X^0}

\def\al{\alpha}

\def\square{{\,\lower0.9pt\vbox{\hrule \hbox{\vrule height 0.2 cm
\hskip 0.2 cm \vrule height 0.2 cm}\hrule}\,}}

\def\lb{\langle}
\def\rb{\rangle}

\def\Afat{\mathbb{A}}
\def\Bfat{\mathbb{B}}
\def\Pfat{\mathbb{P}}

\begin{document}

\begin{titlepage}

\vskip 2em
\begin{center}
{\Large \bf Entropy of temporal entanglement } \\[3em]

\vskip 0.5cm

{\bf
Leonardo Castellani}
\medskip

\vskip 0.5cm

{\sl Dipartimento di Scienze e Innovazione Tecnologica
\\Universit\`a del Piemonte Orientale, viale T. Michel 11, 15121 Alessandria, Italy\\ [.5em] INFN, Sezione di 
Torino, via P. Giuria 1, 10125 Torino, Italy\\ [.5em]
Arnold-Regge Center, via P. Giuria 1, 10125 Torino, Italy
}\\ [4em]
\end{center}

\begin{abstract}
\sk

A recently proposed history formalism is used to define temporal entanglement in quantum systems, and compute its entropy. The procedure is based on the time-reduction of the history density operator,
and allows a symmetrical treatment of space and time correlations. Temporal entanglement entropy
is explicitly calculated in two simple quantum computation circuits.

\end{abstract}

\vskip 10cm\
 \noi \hrule \vskip .2cm \noi {\small
leonardo.castellani@uniupo.it}

\end{titlepage}

\newpage
\setcounter{page}{1}

\tableofcontents

\sect{Introduction}

There are by now a number of proposals for defining and characterizing temporal entanglement \cite{FJV,ZPTGVF,LG,PM,BTCV,HHPBS,Marlettoetal}.  Using the history formalism developed in \cite{LC1,LC2}, we introduce in this note a time-reduced history density matrix. This tool allows for a symmetrical treatment of spatial and temporal entanglement, much in the spirit of the approach of \cite{FJV,ZPTGVF,Marlettoetal}, but within a different framework to describe quantum states over time. 

 Since the work of Feynman \cite{Feynman,FH} (see also Dirac \cite{Dirac}), there have been various formulations of quantum mechanics based on histories, rather than on states at a given time. A very partial list of references, relevant for the present paper, is given in \cite{histories1} - \cite{histories15}.
 
 Here we use the history vector formalism introduced in \cite{LC2}, leading to a
 simple definition of history density operator for a quantum system. Taking ``space" or ``time" partial traces of this operator yields reduced density operators, and these can be used to characterize space or time entanglement between subsystems.
 
 We summarize the formalism in Section 2. In Section 3 the space-reduced 
 history density operator is recalled, and in Section 4 we introduce its time-reduced analogue. The corresponding von Neumann entropy, discussed in Section 5, can be used to detect time correlations. In Section 6 we derive temporal entanglement entropies in two examples
 taken from quantum computation circuits. Section 7 contains some conclusions.
 
\sect{History vector formalism}

\subsection{History vector}

\noi A quantum system over time, together with measuring devices that can be activated at times $t_1,...t_n$, is described by a {\sl history vector} living in $n$-tensor space $\cal H \odot...\odot \cal H$:
\eq
|\Psi\rb = \sum_\al A(\psi,\al) |\al_1\rb \odot ... \odot |\al_n\rb  \label{historyvector}
\en
where $\al = \al_1,...\al_n$ is a sequence of possible measurement results (a ``history"), obtained at times $t_1,...t_n$, and $|\al_i\rb$ are a basis of orthonormal vectors for $\cal H$ at each time $t_i$. If the $\al_i$ eigenvalues are nondegenerate, $|\al_i\rb$ are just the eigenvectors of the observable(s) measured at time $t_i$. For simplicity we assume here nondegenerate eigenvalues (for the general case see \cite{LC2}). The product  $\odot$ has all the properties of a tensor product. The coefficients $A(\psi,\al)$ are the {\sl history amplitudes}, computed as
\eq
A(\psi,\al) = \lb \al_n| U(t_n,t_{n-1}) ~ P_{\alpha_{n-1}} ~ U(t_{n-1},t_{n-2})  \cdots P_{\alpha_{1}}~
U(t_1,t_0)  |\psi\rb \label{amplitude}
\en 
 with $|\psi\rb$ = initial state (at $t_0$). $P_{\alpha_i}$ is the projector on the eigensubspace of $\alpha_i$, and $U(t_{i+1},t_i)$ is the evolution operator between times $t_i$ and $t_{i+1}$.  
 
\noi  The data entering the history vector (\ref{historyvector}) are therefore:
 
\noi - system data: evolution operator (or Hamiltonian), initial state $|\psi\rb$.
 
\noi  - measuring apparatus data: which observables are measured at different times $t_i$.

\subsection{Probabilities}

\noi  Using standard Born rules, it is straightforward to prove that the joint probability $p(\psi,\al)$ of obtaining the sequence $\al_1,...\al_n$ in measurements at times $t_1,...t_n$ is given by 
  the square modulus of the amplitude $A(\psi,\al)$.  If one defines the history projector
  \eq
\mathbb{P}_\al = |\al_1\rb\lb\al_1| \odot ... \odot  |\al_n\rb\lb\al_n|  \label{projalpha}
\en
the familiar formula holds
\eq
p(\psi,\al)= \lb \Psi | \mathbb{P}_\al |\Psi\rb = |A(\psi,\al)|^2   \label{probvector}
\en
generalizing Born rule to measurement sequences. The probabilities $p(\psi,\al)$ satisfy
\eq
\sum_\al p(\psi,\al)= \sum_\al  |A(\psi,\al)|^2 = 1
\en
due to completeness relations for the projectors $P_{\al_i}$ and unitarity of the evolution operators.
As a consequence the history vector is normalized:
\eq
\lb \Psi | \Psi \rb= \sum_\al |A(\psi,\al)|^2 = 1
\en
\noi Defining the chain operator:
\eq
C_{\psi,\al} = P_{\alpha_{n}}  U(t_n,t_{n-1}) ~ P_{\alpha_{n-1}} ~ U(t_{n-1},t_{n-2})  \cdots P_{\alpha_{1}}~
U(t_1,t_0) P_{\psi} \label{chain}
\en
sequence probabilities can also be expressed as
\eq
p(\psi, \al) = Tr (C_{\psi,\al} C_{\psi,\al}^\dagger)
\en

\subsection{Sum rules}

\noi Note that 
\eq
\sum_{\al_{n}} p(\psi, \al_1,\al_2,\cdots  ,\al_{n})  =  p(\psi, \al_1, \al_2, \cdots,  \al_{n-1}) \label{sumrule0}
\en
However other standard sum rules for probabilities are not satisfied in general. For example relations of the type
\eq
\sum_{\al_2} p(\psi, \al_1,\al_2,\al_3)  =  p(\psi, \al_1, \al_3) \label{sumrule}
\en
hold only if the so-called {\it decoherence condition} is satisfied:
\eq
Tr (C_{\psi,\al} C_{\psi,\be}^\dagger) + c.c.= 0~~when~\al \not= \be  \label{decocondition}
\en
as can be checked on the example (\ref{sumrule}) written in terms of chain operators, and easily
generalized. If all the histories we consider are such that the decoherence condition holds, they are said to form a {\it consistent} set \cite{histories1}, and can be assigned probabilities satisfying all the standard sum rules.

In general, histories do not form a consistent set: interference effects between them can be important, as in the case of the double slit experiment. For this reason we do not limit ourselves to consistent sets. Formula (\ref{probvector}) for the probability of successive measurement outcomes holds true in any case.

\subsection{Scalar and tensor products in history space}

Scalar and tensor products in history space, i.e. the vector space spanned by the basis vectors 
$ |\al_1\rb \odot ... \odot |\al_n\rb$, can be defined as in ordinary tensor spaces.
\sk
\noi {\sl Scalar product:}
\eq
 (\lb \al_1| \odot ... \odot \lb \al_n|) ( |\be_1\rb \odot ... \odot |\be_n\rb) \equiv
  \lb \al_1 |\be_1 \rb ... \lb \al_n  | \be_n\rb
  \en
and extended by (anti)linearity on all linear combinations of these vectors. This also defines
bra vectors in history space.

\sk
\noi {\sl Tensor product: }
\eq
 (|\al_1\rb \odot ... \odot |\al_n\rb) ( |\be_1\rb \odot ... \odot |\be_n\rb) \equiv
  |\al_1\rb |\be_1 \rb \odot ... \odot |\al_n \rb | \be_n\rb
  \en
and extended by bilinearity on all linear combinations of these vectors. No symbol is used 
for this tensor product, to distinguish it from the tensor product $\odot$ involving different times $t_k$.
\sk
This tensor product allows a definition of {\sl product history states}, which are defined to be
expressible in the form:
\eq
 (\sum_\al A(\phi,\al) |\al_1\rb \odot ... \odot |\al_n\rb ) (  \sum_\be A(\chi,\be) |\be_1\rb \odot ... \odot |\be_n\rb ) \label{hproduct}
 \en
or, using bilinearity:
\eq
\sum_{\al,\be} A(\phi,\al) A(\chi,\be) |\al_1 \be_1 \rb \odot...\odot |\al_n \be_n \rb
\en
with $|\al_i \be_i \rb \equiv |\al_i \rb | \be_i \rb$ for short. A product history state is thus
characterized by factorized amplitudes $A(\psi,\al,\be) =  A(\phi,\al) A(\chi,\be)$. If the history state cannot be expressed as a product, 
we define it to be {\sl history entangled}. In this case,
results of measurements on system A are correlated with those on system B and viceversa.

\subsection{History density matrix}

A system in the history state $|\Psi\rb$ can be described by the {\sl history density matrix}:
\eq
\rho = |\Psi\rb \lb \Psi |
\en
a positive operator satisfying $Tr(\rho) = 1$ (due to $\lb \Psi |\Psi \rb = 1$). A mixed history state has density matrix
\eq
\rho = \sum_i p_i  |\Psi_i\rb \lb \Psi_i |
\en
with $\sum_i p_i = 1$, and $\{ |\Psi_i \rb \}$ an ensemble of history states. Probabilities of measuring sequences $\al = \al_1,...\al_n$ in history state $\rho$ are given by the standard formula:
\eq
p(\al_1,...\al_n) = Tr(\rho ~ \mathbb{P}_\al)
\en
cf. equation (\ref{probvector}) for pure states.

\sect{Space-reduced density matrix}

\noi Consider now a system AB composed by two subsystems A and B, and devices measuring
observables $\Afat_i = A_i \otimes I$ and $\Bfat_i=I \otimes B_i$  at each $t_i$. Its history state is
\eq
|\Psi^{AB}\rb = \sum_{\al,\be} A(\psi,\al,\be) |\al_1 \be_1 \rb \odot...\odot |\al_n \be_n \rb
\label{PsiAB}
\en
where $\al_i,\be_i$ are the possible outcomes of a joint measurement at time $t_i$ of 
$\Afat_i$ and $\Bfat_i$.  The amplitudes $A(\psi,\al,\be)$ are computed using the general formula
(\ref{amplitude}), with projectors 
\eq
\Pfat_{\al_i,\be_i} = |\al_i,\be_i \rb\lb \al_i,\be_i|= |\al_i \rb\lb \al_i| \otimes |\be_i \rb\lb\be_i|
\en
corresponding to the eigenvalues $\al_i,\be_i$. The density matrix of AB is
\eqa
& & \rho^{AB} = |\Psi^{AB}\rb \lb \Psi^{AB}| = \nonumber \\
 & & ~~~~~ = \sum_{\al,\be,{\al}', {\be}'} 
A(\psi,\al,\be) A(\psi,{\al}',{\be}')^* (|\al_1 \be_1 \rb \odot...\odot |\al_n \be_n \rb )
 (\lb \al'_1 \be'_1 | \odot...\odot \lb \al'_n \be'_n |) \nonumber\\
\label{rhoAB}
\ena
We define {\sl space-reduced density matrices} by partially tracing on the subsystems:
\eq
\rho^A \equiv Tr_B  ( \rho^{AB} ), ~~~\rho^B \equiv Tr_A  ( \rho^{AB} ) \label{rhoA}
\en
In general $\rho^A$ and $\rho^B$ will not describe pure history states anymore. These reduced density matrices can be used to compute statistics for
measurement sequences on the subsystems.  Taking for example the partial trace on B of (\ref{rhoAB}) yields:
\eq
\rho^A = \sum_{\alpha,{\alpha}',\beta} A(\psi,\alpha,\beta) A^* (\psi,{\alpha}',\beta) 
(|\al_1  \rb \odot...\odot |\al_n \rb) ( \lb {\al}'_1| \odot...\odot \lb {\al}'_n  |),
\en
a positive operator with unit trace. The standard expression in terms of $\rho^A$ for Alice's probability to obtain the sequence  $\al$ is 
\eq
p(\alpha) = Tr(\rho^A \Pfat_\alpha)  \label{prho}
\en
with
\eq
\Pfat_\alpha = (P_{\al_1} \otimes I) \odot \cdots \odot (P_{\al_n} \otimes I),~~~P_{\al_i} =
 |\al_i\rb\lb\al_i| \label{Pfat}
\en
The prescription (\ref{prho}) yields
\eq
p(\alpha) =  \sum_\beta  |A(\psi,\al,\be) |^2 = \sum_\beta p(\alpha,\beta)  \label{psum}
\en
i.e. the probability for Alice to obtain the sequence $\al$ in measuring the observables $A_i$.
\sk
On the other hand,  the probability for Alice to obtain the sequence
$\al_1,...\al_n$  {\sl with no measurements on Bob's part} is in general different from (\ref{psum}).
Indeed, the history vector of the composite system is different, since only
Alice's measuring device is activated, and reads
\eq
|\Psi^{AB}\rb = \sum_\al A(\psi,\al) |\al_1  \rb \odot...\odot |\al_n \rb \label{PsiAB2}
\en
where the amplitudes $A(\psi,\al)$ are obtained from the general formula (\ref{amplitude})
using the projectors $P_{\al_i}$ of (\ref{Pfat}). Here the reduced density operator $\rho^A$ is simply
\eq
\rho^A =  \sum_{\al,{\al}'}  A(\psi,\al) A(\psi,{\al}')^*  |\al_1  \rb \odot...\odot |\al_n \rb 
 \lb \al'_1  | \odot...\odot \lb \al'_n |
 \en
 (the trace on B has no effect, since history vectors contain only results of Alice), and the probability of Alice finding the sequence $\al$ is
 \eq
 p(\al) = Tr(\rho^A \Pfat_\alpha) =  |A(\psi,\al)|^2 \label{palpha}
 \en
 differing in general from (\ref{psum}). Indeed $\sum_\beta A(\psi,\alpha,\beta) = A(\psi,\alpha)$ because of 
 the completeness relation (at each time $t_i$)  $\sum_\beta |\beta_i\rb\lb \beta_i| =I$, so that
 \eq
 p(\alpha) = |A(\psi,\al)|^2 =  | \sum_\beta A(\psi,\al,\beta)|^2 \label{palpha2}
 \en
 differing, in general, from (\ref{psum}).

 In fact the probabilities (\ref{psum}) and (\ref{palpha2}) coincide only when the evolution operator is factorized $U=U^A \otimes U^B$, i.e. when A and B do not interact \cite{LC2}.
Thus, if there is no interaction, Bob cannot communicate with Alice by activating (or not activating) his measuring devices.

\sect{Time-reduced density matrix}

Partial traces of the history density matrix can be taken also on the Hilbert spaces ${\cal H}_i$ corresponding to different times $t_{\{k\}} = t_{k_1},...t_{k_p}$, $p<n$. We call the resulting density matrices, involving only the complementary times $t_{\{j\}} = t_{j_1},...t_{j_m}$ (i.e. with $j_1,...j_m$ and $k_1,...k_p$ having no intersection, and union coinciding with $1,...n$), {\sl time-reduced density matrices}. They 
are used to compute sequence probabilities corresponding to measurements at times $t_{\{j\}} $, given that measurements are performed also at times $t_{\{k\}}$ without registering their result. Thus they describe statistics for an experimenter that has access only to the measuring apparati at times $t_{\{j\}}$, while
the system gets measured at all times $t_i=t_1,...t_n$.
\sk
\noi Consider a system described by the (pure) history vector (\ref{historyvector}). Its density matrix is
\eq
\rho =  \sum_{\al,{\al}'}  A(\psi,\al) A(\psi,{\al}')^*  |\al_1  \rb \odot...\odot |\al_n \rb 
 \lb \al'_1  | \odot...\odot \lb \al'_n |
\en
Dividing the set $\al = \al_1,...\al_n$ into the complementary sets $\al_{\{j \}} = \al_{j_1},...\al_{i_m}$ and
$\al_{\{k \}} = \al_{k_1},...\al_{k_p}$, the $\{j \}$-time reduced density matrix is defined by
\eq
\rho^{\{j \}} =Tr_{\{k \}} \rho = \sum_{\al_{\{j\}},\al'_{\{j\}}} \sum_{\al_{\{k\}}} A(\psi, \al_{\{j\}},\al_{\{k\}})
A^*(\psi, \al'_{\{j\}},\al_{\{k\}})   | \al_{\{j\}} \rb \lb \al'_{\{j\}} |
\en
with $| \al_{\{j\}} \rb \equiv  |\al_{j_1}  \rb \odot...\odot |\al_{j_m} \rb $.
Using the standard formula we find the probability for the sequence $\al_{\{j \}}$
\eq
p(\al_{\{j \}}) = Tr (\Pfat_{\al_{\{j \}} } \rho^{\{j \}} ) = \sum_{\al_{\{k\}}} |A(\psi, \al_{\{j\}},\al_{\{k\}})|^2 \label{probalj}
\en
where $\Pfat_{\al_{\{j \}} }$ is the projector on the (sub)history $\al_{\{j \}}$, given by eq. (\ref{projalpha}) with the $\alpha$'s in $\al_{\{j \}}$.

On the other hand, if no measurements are performed at complementary times $t_{\{k\}}$, the probability for the same sequence $\al_{\{j \}}$ is simply
\eq
p(\al_{\{j \}}) = |A(\psi, \al_{\{j\}}|^2
\en
in general differing from (\ref{probalj}), see the discussion on sum rules after (\ref{sumrule0}). Can this difference be used to violate causality ? More precisely, can a future measurement by Alice be detected by herself in the past ? The answer is of course negative, but the formal reason is interesting. It is based on the marginal rules of Section 2.3:  no difference arises in the probabilities for an experimenter having access to measurement results up to time $t$, whether the system gets measured or not at times $t' >t$, due to the validity of formula (\ref{sumrule0}) that reproduces a classical sum rule. When $t' < t$ this formula does not hold, and indeed
past measurements have a verifiable impact on present statistics. This asymmetry in time is entirely due to the particular marginal rules for quantum probabilities of sequences.

\sect{Temporal entanglement}

\subsection{Time tensor product between histories}

The ``time" tensor product $\odot$ introduced in Section 2.1 can be extended to a time tensor product between histories, in contradistinction with the product defined in Section 2.4, which could be referred to as a ``space" tensor product.
\sk
The definition is given by the merging rule:
\eq
| \al_{\{j\}} \rb \odot | \al_{\{k\}} \rb \equiv | \al_{\{i\}} \rb
\en
with ${\{j\}}$ and ${\{k\}}$ having no intersection and union equal to ${\{i\}} $. For example
\eq
( |\al_{1}  \rb \odot  |\al_{3}  \rb \odot  |\al_{5}  \rb) \odot ( |\al_{2}  \rb \odot  |\al_{6}  \rb ) =
 |\al_{1}  \rb \odot  |\al_{2}  \rb \odot  |\al_{3}  \rb \odot  |\al_{5}  \rb \odot  |\al_{6}  \rb
 \en

We then denote $|\Psi\rb$  as a {\sl time-separable history state} if it can be expressed as a time product 
of two history states:
\eq
|\Psi\rb = |\Psi_1 \rb \odot |\Psi_2\rb
\en
in analogy with the ``space" product history state of Section 2.4. 
Similarly, we find here that a history state 
\eq
|\Psi \rb =  \sum_{\al_{\{i\}}} A(\psi, \al_{\{i\}}) | \al_{\{i\}} \rb
\en
is time-separable if and only if the amplitudes factorize
\eq
A(\psi, \al_{\{i\}}) = A(\psi, \al_{\{j\}}) A(\psi, \al_{\{k\}})
\en
with ${\{j\}}$ and ${\{k\}}$ having no intersection and union equal to ${\{i\}} $.
\sk
As a consequence probabilities factorize, and there are no temporal correlations between 
measurement results $ \al_{\{j\}}$ and $ \al_{\{k\}}$. If the amplitudes do not factorize, we call
the history state a {\sl temporally entangled} history state. Note that a time-separable state can still contain
entangled sub-histories, exactly as a (space) separable state in a composite system AB can still be entangled within the subsystems A and B.

\subsection{Temporal entanglement entropy}

History entropy has been defined in \cite{LC2} as the von Neumann entropy associated to the history state $\rho$:
\eq
S(\rho) = - Tr (\rho \log \rho)
\en
We have seen in \cite{LC2} that when $\rho$ describes a (pure) space entangled system, partial traces of $\rho$ describe mixed history states. The same happens for (pure) time entangled systems: partial time-traces yield reduced density matrices describing mixed states. Examples taken from quantum computation circuits are discussed in the next Section.
\sk
We call {\sl temporal entanglement entropy} the von Neumann entropy corresponding to the time-reduced density matrix.

\sect{Examples}

In this Section we examine two examples of quantum systems evolving from a given initial state, and subjected to successive measurements.
They are taken from simple quantum computation circuits, where unitary gates determine the evolution between measurements. 
Only two gates are used: the Hadamard one-qubit gate $H$ defined by:
\eq
H |0\rb = {1\over \sqrt{2}} (|0\rb + |1\rb),~~~H |1\rb = {1\over \sqrt{2}} (|0\rb - |1\rb)
\en
and the two-qubit $CNOT$ gate:
\eq
CNOT |00\rb = |00\rb,~CNOT |01\rb = |01\rb,~CNOT |10\rb = |11\rb,~CNOT |11\rb = |10\rb
\en

\subsection{Entangler}

The two-qubit entangler circuit of Fig. 1, with initial state $|00\rb$, produces the entangled state 
$|\chi\rb= {1 \over \sqrt{2} }(|00\rb+|11\rb)$:

\includegraphics[scale=0.45]{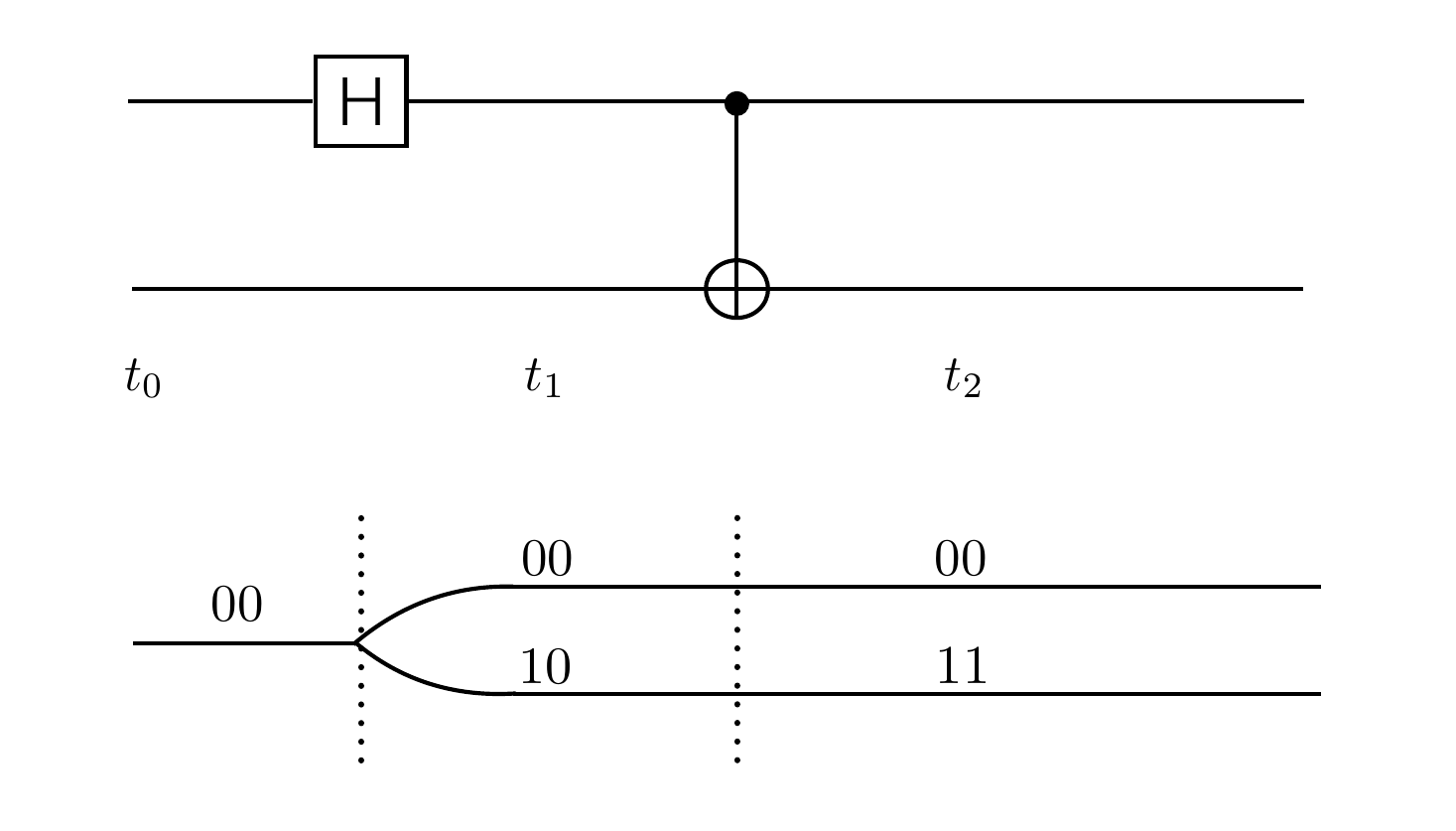}

\noi {\bf Fig. 1 } {\small Entangler circuit, and history graph for initial state $|00\rb$}
\sk

 The history state that describes the system together with its
measuring devices\footnote{we consider here measurements in the computational basis.} at times $t_1$ and $t_2$ is 
\eq
|\Psi\rb = A(00,00,00) |00\rb \odot |00\rb +  A(00,10,11) |10\rb \odot |11\rb
\en
where
\eqa
& & A(00,00,00)=\lb 00|CNOT|00\rb\lb 00|H \otimes I |00\rb = {1 \over \sqrt{2}} \\
& & A(00,10,11)=\lb 11|CNOT|10\rb\lb 10|H \otimes I |00\rb = {1 \over \sqrt{2}} 
\ena
are the only nonvanishing amplitudes. This simple system exhibits both space and time entanglement.
Space entanglement is due to (ordinary) entanglement in the final state at $t_2$. Time entanglement
is due to temporal correlations: the outcomes of measurements at $t_1$ are correlated with the outcomes at $t_2$. In other words, the history amplitudes $A(00, \al_1\be_1,\al_2\be_2)$ do not space-factorize as $A(0, \al_1, \al_2) A(0, \be_1, \be_2)$, and do not time-factorize as
$A(00, \al_1\be_1) A(00,\al_2\be_2)$. 
\sk
\noi The density matrix is given by
\eq
\rho = {1 \over 2} (|00\rb \odot |00\rb + |10\rb \odot |11\rb )(\lb 00\| \odot \lb 00| + \lb 10| \odot \lb 11|) 
\en
The space-reduced density matrices are
\eqa
& & \rho^A = Tr_B (\rho) = {1 \over 2} [(|0\rb \odot |0\rb)   (\lb 0| \odot  \lb 0|)+( |1\rb  \odot  |1\rb)(\lb 1| \odot \lb 1|)] \\
& & \rho^B = Tr_A (\rho) = {1 \over 2} [(|0\rb \odot |0\rb)   (\lb 0| \odot  \lb 0|)+( |0\rb  \odot  |1\rb)(\lb 0| \odot \lb 1|)]
\ena
i.e. mixed history states, to be expected since $\rho$ is a pure space-entangled history state\footnote{These quantum mixtures would be called, in D' Espagnat's \cite{dEspagnat} terms, ``improper" mixtures. See however \cite{LC3,LC4} for a critique on the distinction between proper and improper mixtures.}

\noi The time reduced density matrices are
\eqa
& & \rho^{\{1\}} = Tr_{\{2 \}} \rho = {1 \over 2} (|00\rb\lb 00| + |10\rb\lb 10|) \\
& & \rho^{\{2\}} = Tr_{\{1 \}} \rho = {1 \over 2} (|00\rb\lb 00| + |11\rb\lb 11|)
\ena
i.e. mixed history states (in this case states corresponding to a single time), to be expected since $\rho$ is a pure time-entangled history state.
\sk
The history entropy corresponding to $\rho$ is $S(\rho)=0$ since $\rho$ is a pure history state, 
while the space and time entanglement entropies are $S(\rho^A)=S(\rho^B)=1$ and
$S(\rho^{\{1\}})=S(\rho^{\{2\}})=1$ since they all have two eigenvalues equal to ${1 \over 2}$.

\subsection{Teleportation}

The teleportation circuit \cite{teleportation}  is the three-qubit circuit given in Fig. 2, where the upper two qubits belong to Alice, and the lower one to Bob.
\sk
\includegraphics[scale=0.45]{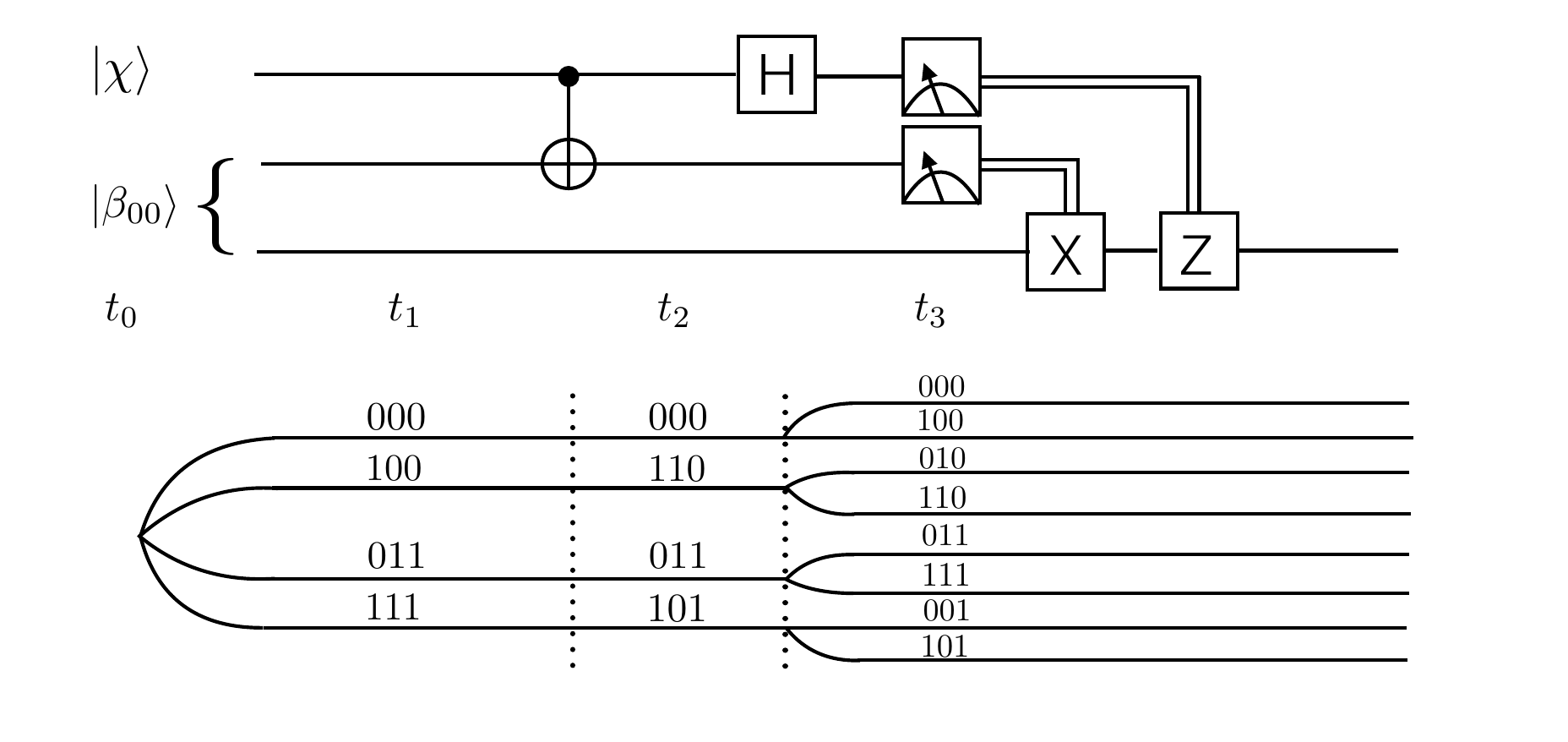}

\noi {\bf Fig. 2 } {\small Teleportation circuit, and history graph for initial state $(\al |0\rb + \be |1\rb) \otimes   {1 \over \sqrt{2}} (|00 \rb + |11 \rb)$}
\sk
\noi The initial state is a three-qubit state, tensor product of the single qubit $|\chi\rb = \al |0\rb + \be |1\rb$ to be teleported and the 2-qubit entangled Bell state
$|\be_{00}\rb = {1 \over \sqrt{2}} (|00 \rb + |11 \rb)$. The history vector contains
8 histories:
\eqa
& & |\Psi\rb = {1 \over 2} (\al |000\rb \odot |000\rb \odot  |000\rb  + \al|000\rb \odot |000\rb \odot  |100\rb  + \nonumber \\
& & ~~~~~~~~+ \be |100\rb \odot |110\rb \odot  |010\rb - \be |100\rb \odot |110\rb \odot  |110\rb \nonumber \\
& & ~~~~~~~~+\al |011\rb \odot |011\rb \odot  |011\rb + \al |011\rb \odot |011\rb \odot  |111\rb \nonumber \\
& & ~~~~~~~~+\be |111\rb \odot |101\rb \odot  |001\rb -\be |111\rb \odot |101\rb \odot  |101\rb \nonumber \\
\label{historytel}
\ena
the amplitudes being given by 
\eq
A(\chi \otimes \be_{00},\al_1,\al_2,\al_3) = \lb \al_3 | H_1 P_{\al_2} {\rm CNOT}_{1,2} P_{\al_1} |\chi \otimes \be_{00} \rb 
\en
For example
\eq
A(\chi \otimes \be_{00},000,000,000) = \lb 000|  H_1  |000\rb\lb 000|  {\rm CNOT}_{1,2}  |000\rb\lb 000|  \chi \otimes \be_{00} \rb = \al / 2
\en
where $H_1 \equiv H \otimes I \otimes I$ and $ {\rm CNOT_{1,2}} \equiv {\rm CNOT} \otimes I$.  
\sk
\noi As in the entangler example, here too the history state is space and time entangled.
The partial traces of the history density matrix $\rho = |\Psi\rb\lb\Psi|$ yield the density
matrix for Alice:
\eq
\rho^A = Tr_B (\rho)={1\over 2} |\phi\rb\lb \phi| + {1\over 2} |\chi\rb\lb \chi| 
\en
with
\eqa
|\phi\rb &=& {1 \over \sqrt{2}}(\alpha |00\rb\odot|00\rb\odot|00\rb -\alpha |00\rb\odot|00\rb\odot|10\rb
+\beta  |10\rb\odot|11\rb\odot|01\rb - \beta  |10\rb\odot|11\rb\odot|11\rb) \nonumber \\
 |\chi\rb &=& {1 \over \sqrt{2}}(\alpha |01\rb\odot|01\rb\odot|01\rb -\alpha |01\rb\odot|01\rb\odot|11\rb
+\beta  |11\rb\odot|10\rb\odot|00\rb - \beta  |11\rb\odot|10\rb\odot|10\rb) \nonumber \\
\ena
and the density matrix for Bob:
\eqa
 \rho^B = Tr_A (\rho) &=&{1\over 2}(|0\rb\odot|0\rb\odot|0\rb \lb 0|\odot\lb 0|\odot \lb 0|+
|1\rb\odot|1\rb\odot|1\rb \lb 1|\odot\lb 1|\odot \lb 1|) \nonumber \\ \label{rhoB}
\ena
As expected, both reduced history density operators describe mixed states. They both have two nonzero eigenvalues equal to $1/2$, and the (space) entanglement entropy is therefore
$ S(\rho^A) = S(\rho^B) = - {1 \over 2} \log {1 \over 2} - {1 \over 2} \log {1 \over 2} = 1$

Next we compute the time-reduced density matrices. We can take partial traces of $\rho$ over any combination of $t_1,t_2,t_3$. 
For example taking the partial trace over $t_1$ and $t_2$ yields the time reduced density matrix for the system at time $t_3$:
\eq
\rho^{\{3\}} = Tr_{t_1,t_2} (\rho) = {|\alpha|^2 \over 2} ( [+00] + [+11]) + {|\beta|^2 \over 2} ( [-10] + [-01])
\en
where $[\pm 00]$ indicates the projector on the vector ${1 \over \sqrt{2}} (|0\rb \pm |1\rb) |00\rb$, etc.
This density matrix describes a mixed state. Its eigenvalues are ${|\alpha|^2 \over 2} ,{|\alpha|^2 \over 2} ,{|\beta|^2 \over 2} ,{|\beta|^2 \over 2} $, and therefore the time entanglement entropy is
\eq
S(\rho^{\{3\}} ) =  -|\al|^2 \log {|\al|^2 } - |\be|^2 \log {|\be|^2}+1 \label{Srho3}
\en
Setting $p=|\al|^2$, the entropy $S(p)=1-p\log p-(1-p) \log (1-p)$ is maximum and equal to 
$\log 2+1=2$ when $p=1/2$, and is minimum and equal to $1$ when $p=0,1$.
\sk
Taking the partial trace on the time complementary to $t_1,t_2$, i.e. on $t_3$, yields the
time reduced density matrix:
\eq
\rho^{\{1,2\}} = Tr_{t_3} (\rho) = {|\alpha|^2 \over 2} ( [000 \odot 000] + [011 \odot 011]) + {|\beta|^2 \over 2} ( [100 \odot 110] + [111 \odot 101])
\en
where $[000 \odot 000]$ is the projector on the history vector  $|000\rb \odot |000\rb$ etc. This reduced density matrix has, as expected, the same eigenvalues as $\rho^{\{3\}} $, and corresponds therefore to the same  time entanglement entropy.
\sk
Finally, taking the partial trace of $\rho$ on $t_2$, $t_3$ yields the time-reduced history density matrix:
\eq
\rho^{\{1\}} = Tr_{t_2,t_3} (\rho)= {|\alpha|^2 \over 2} ( [000] + [011]) + {|\beta|^2 \over 2} ( [100] + [111])
\label{rho1}
\en
corresponding again to the same time entanglement entropy as in (\ref{Srho3}). Note that if no measurements are performed at $t_2,t_3$ the history density matrix at $t_1$ remains the same as in 
(\ref{rho1}), due to quantum marginal rules of type (\ref{sumrule0}). Indeed performing or not measurements at $t>t_1$ cannot change statistics at $t_1$.

\sect{Conclusions}

The history formalism of \cite{LC2} permits a symmetrical treatment of space and time correlations,
based on the reduced history density operator.  The same history density $\rho$ can be partially traced both on space and time subsystems: in fact space and time partial tracings commute, so that
the resulting reduced density does not depend on the order of the tracings, and describes the statistics of an observer having limited (in space and time) access to the system.

\section*{Acknowledgements}

This work is supported by the research funds of the Eastern Piedmont University and INFN - Torino Section.

\vfill\eject
\end{document}